# Exploring the Privacy Concerns in Permissionless Blockchain Networks and Potential Solutions

Talgar Bayan
Department of Computer Science
University of Manchester
Manchester, UK
talgar.bayan@manchester.ac.uk
ORCID ID:0000-0001-8783-2229

Richard Banach
Department of Computer Science
University of Manchester
Manchester, UK
richard.banach@manchester.ac.uk
ORCID ID:0000-0002-0243-9434

*Abstract*— In recent years, permissionless blockchains have gained significant attention for their ability to secure and provide transparency in transactions. The development of blockchain technology has shifted from cryptocurrency to decentralized finance, benefiting millions of unbanked individuals, and serving as the foundation of Web3, which aims to provide the next generation of the internet with data ownership for users. The rise of NFTs has also helped artists and creative workers to protect their intellectual property and reap the benefits of their work. However, privacy risks associated with permissionless blockchains have become a major concern for individuals and institutions. The role of blockchain in the transition from Web2 to Web3 is crucial, as it is rapidly evolving. As more individuals, institutions, and organizations adopt this technology, it becomes increasingly important to closely monitor the new risks associated with permissionless blockchains and provide updated solutions to mitigate them.

This paper endeavors to examine the privacy risks inherent in permissionless blockchains, including Remote Procedure Call (RPC) issues, Ethereum Name Service (ENS), miner extractable value (MEV) bots, on-chain data analysis, data breaches, transaction linking, transaction metadata, and others. The existing solutions to these privacy risks, such as zero-knowledge proofs, ring signatures, Hyperledger Fabric, and stealth addresses, shall be analyzed. Finally, suggestions for the future improvement of privacy solutions in the permissionless blockchain space shall be put forward.

## I. Introduction

A permissionless blockchain is a decentralized, distributed ledger that enables secure, transparent, and tamper-proof transactions. It operates on a peer-to-peer network, where all participants have equal access to the network and its data and can participate in the validation of transactions. As these blockchains are open source, anyone can join the network and validate transactions. The use of a public ledger, accessible to all participants, increases accountability, improves security, and enhances trust. However, this transparency also raises privacy concerns, as digital assets, personal information, and financial transactions of participants can be easily viewed by anyone, potentially leading to hacking and theft.

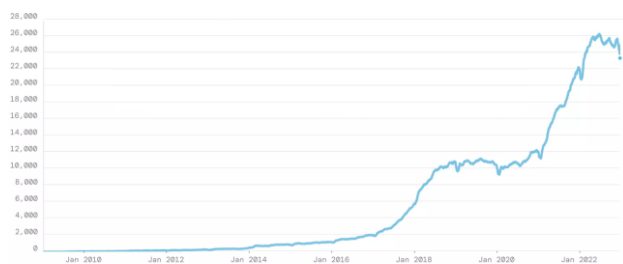

Figure 1. Monthly active developers since 2009.

We conducted a comprehensive analysis of the leading permissionless blockchains, focusing on those with the highest levels of user engagement and active developer (shown in Figure 1) communities. The results of our analysis are presented in Table 1. Over the past seven years, the permissionless blockchain sector has seen tremendous growth and innovation, with a notable increase in the number of monthly active developers, now numbering over 22,000 [2].

TABLE I. POPULAR PERMISSIONLESS BLOCKCHAINS

| Blockchain | Type | TVL | Transactions (daily) | Protocols | Monthly Active Developers |
|---|---|---|---|---|---|
| Ethereum | Layer 1 | 27B [3] | 1.2M [4] | 693 [3] | 5734 [2] |
| BSC | Layer 1 | 4.9B [3] | 3.1M [3] | 540 [6] | 150 [2] |
| Solana | Layer 1 | 245M [3] | 0.13M [7] | 93 [3] | 2200 [2] |
| Polygon | Layer 2 | 1.17B [3] | 3M [5] | 367 [3] | 1100 [2] |
| Cosmos | Layer 1 | 1B [3] | N/A | 281 [3] | 1500 [2] |
| BTC | Layer 1 | 131M [3] | 300092 [3] | 4 [3] | 920 [2] |
| Near | Layer 1 | 81M [3] | 374268 [8] | 15 [3] | 830 [2] |

The rapid adoption of permissionless blockchain technology has resulted in a surge of interest from various industries and even countries. With the growth of capital, start-ups, organizations, and nations embracing blockchain, the potential risks associated with public blockchains have become increasingly pressing. Despite the transparency benefits offered by blockchain, the openness of the network also presents privacy concerns. The recent FTX incident highlights the critical need to address the privacy risks inherent in on-chain data. It is imperative to consider and address these challenges to ensure the continued growth and success of the blockchain industry.

This paper delves into the ongoing conflict between transparency and privacy in permissionless blockchains. We aim to provide a comprehensive analysis of the latest privacy risks that have arisen in the sector, from MEV bots to smart contract vulnerabilities. The main objective of this paper is twofold:

- To examine and highlight the most pressing privacy concerns in the permissionless blockchain space, including risks posed by MEV bots, RPC, name service providers (such as ENS), and the latest smart contract risks.
- To propose potential solutions to these privacy risks in the form of Zero Knowledge Proofs, zkEVM blockchains, ring signature technology, and the implementation of modular blockchains like Hyperledger Fabric. By doing so, we aim to provide valuable insights and recommendations to help mitigate privacy risks in the rapidly growing permissionless blockchain industry.

## II. RELATED WORKS

This section summarizes the related works on privacy preservation in permissionless blockchain networks and some alternative solutions for that. In [1] Punishment Not Reward (PnR) blockchain architecture was proposed to address the issue of balancing privacy and openness in permissionless blockchain networks such as Ethereum and Bitcoin. The author suggested two methods, denial of service to the application and/or revocation of anonymity, as alternatives to the traditional reward-based approach for block creation to maintain the network's privacy.

In [10] a qualitative comparison was presented for privacy-preserving methods of engineering data, the methods include proxy encryption, homomorphic encryption, ZKP, and a trusted execution environment, the result indicates that approaches that rely on a trusted third party for preserving participant privacy do not provide sufficiently strong guarantees that sensitive data would not be exposed in modern data ecosystems. In [11] proposed a privacy-preserving permissioned blockchain solution using Hyperledger Fabric network with the implementation of ZKP that ensures the patients' data privacy, the Idmix suit was used to preserve features, such as anonymity and unlikability. In [12] a systematic review of the current state of the art on privacy-preserving research solutions and mechanisms in blockchain was presented, it also discussed challenges in blockchain scenarios like postquantum computing resistance, Malicious-Curious TTPS, and regulations. Table II shows the results of comparisons of the most related solutions for permissionless privacy concerns.

TABLE II. RELATED SOLUTIONS COMPARISON

| Related solutions | Multi-signature | ZKP | HLF | Ring signature | Stealth Addresses privacy-enhancing | zkEVM |
|---|---|---|---|---|---|---|
| [12] [15] [9] | | ✓ | | ✓ | ✓ | |
| [1] | | ✓ | | | | ✓ |
| [11] [13] | | | ✓ | | | |
| [18] | | ✓ | | ✓ | | |
| [14] [16] [17] | | ✓ | | | ✓ | |
| [10] | | ✓ | ✓ | | | |
| Our work | ✓ | ✓ | ✓ | ✓ | ✓ | ✓ |

In general, due to the rapid growth of the permissionless blockchain space, including various applications and areas such as web 3, NFT, decentralized applications, and social graphs, etc. Except for the other authors, we have more aspects like multi-sig technique, MEV bot, etc. We could notice that there is a gap between the intensive industrial development and research investigation and argues that it is important to regularly explore and analyze permissionless blockchain networks.

## III. EXPLORING THE PRIVACY CONCERNS IN PERMISSIONLESS BLOCKCHAIN NETWORKS

Blockchain's transparent architecture comes with privacy risks. Previously, these risks were basic and manageable, but with the rapid growth of Defi, NFT, and web3 dApps, new risks have emerged. We shall discuss some of the latest typical risks in this paper.

### A. Risk Associated with Private Key Management

A private key in a public blockchain is a secret digital code used to access and manage a user's assets and transactions. The private key acts as a password that allows users to sign transactions and change their blockchain assets. It is crucial for users to keep their private keys safe and secure, as anyone who has access to a private key can control and manipulate the assets associated with that key. Poor management of private keys can result in serious risks, including:

- Loss of funds: If a user loses or forgets their private key, they would no longer be able to access their assets on the blockchain, effectively losing their funds.
- Hacking or theft: If a user's private key is stolen or obtained by a malicious actor, they can use it to control and manipulate the user's assets and transactions on the blockchain.
- Data breaches: Storing private keys in online environments or sharing them with others can lead to data breaches and potential hacks, compromising the safety and security of users' assets and transactions.

An example of the consequences of poor private key management can be seen in the Mt. Gox incident, where the exchange suffered a major hack that resulted in the loss of 850,000 Bitcoin, worth over $450 million at the time. The hack was later attributed to poor private key management, as the private keys were stored in an online environment and were easily accessible to the attacker.

It is important for users of permissionless blockchain networks to understand the risks associated with private key management and to take steps to keep their private keys safe and secure, such as using hardware wallets or following best practices for private key storage and management.

*B. Transaction Linking and On-chain analysis*

Transaction linking and on-chain analysis in permissionless blockchain pose privacy risks to users by potentially revealing the identity and behavior of users through analysis of publicly visible transaction data. This can expose sensitive information such as financial transactions or personal data. To mitigate these risks, users can employ privacy-enhancing techniques, but these methods are not foolproof and can still be vulnerable to sophisticated analysis techniques. It is important for both users and the wider community to understand these risks and work towards improving the privacy and security of permissionless blockchain technology.

*C. Blockchain Transaction Metadata Risk*

Blockchain transaction metadata can pose privacy risks as it can potentially reveal information about the sender and receiver of a transaction, their behaviors, and other sensitive information. For example, transaction metadata can be used to:
- De-anonymize users: By analyzing transaction patterns and linking transactions to real-world identities, the anonymity of users can be compromised.
- Track the flow of funds: Transaction metadata can be used to trace the flow of funds on a blockchain network, potentially revealing sensitive financial information.
- Identify behavior patterns: Analysis of transaction metadata can reveal behavior patterns and habits of users, such as their spending habits or their preferred blockchain-based services.

These privacy risks are especially pronounced in public blockchain networks where transactions are publicly visible and can be easily analyzed.

*D. Blockchain Name Service and Unique Address Risk*

A blockchain wallet address serves as a unique identifier for transactions on a blockchain network. While these addresses are generated as complex strings, third-party services can provide more user-friendly options such as ENS, which acts as a blockchain domain name. However, this convenience can come with risks as users may lose control of their private keys during the generation process, leading to the loss of assets and potential hacking. Understanding and addressing the challenges of private key management is critical for ensuring the security and protection of blockchain assets.

*E. RPC Privacy Risk*

RPC (Remote Procedure Call) is a protocol that allows a client to make requests to a server over a network. In public blockchain networks, RPC is used to communicate with a node in the network and interact with the blockchain. While RPC provides a convenient and efficient way to interact with a blockchain, it can also pose security risks if not properly secured.

Some of the risks associated with RPC in public blockchain networks include:
- Man-in-the-middle attacks: If the communication between the client and the server is not encrypted, a malicious actor can intercept and manipulate the data being transmitted, potentially compromising the security of the client and the blockchain.
- Unauthorized access: If an attacker can gain access to the RPC interface, they can potentially control and manipulate the blockchain node, compromising the security and integrity of the network.
- Unsecured authentication: If the authentication process for accessing the RPC interface is not secure, an attacker may be able to bypass the authentication and gain unauthorized access to the RPC interface.

To mitigate these risks, it is important to properly secure the RPC interface, such as using encrypted communication and implementing strong authentication measures. Additionally, it is important to follow best practices for securing blockchain nodes and networks, such as regularly updating software and monitoring for security threats.

*F. MEV bot Risk*

MEV (Miner Extractable Value) bots are computer programs that exploit the vulnerabilities in blockchain protocols to extract profits or value from the network. In other words, MEV bots are designed to manipulate the blockchain network and extract profits from transactions or block rewards.

MEV bots can operate in a variety of ways, such as frontrunning transactions, censoring transactions, or manipulating the price of assets. For example, a front running MEV bot might place a trade ahead of a large order, taking advantage of the change in price that the large order would cause. This can result in the MEV bot earning a profit at the expense of other traders.

MEV bots are considered a threat to the integrity of blockchain networks and can have a negative impact on the overall health and stability of the network.

*G. Smart Contract Risk*

Smart contracts provide a valuable tool for automating and enforcing contractual agreements on a blockchain network. However, the transparency and public accessibility of permissionless blockchains can also lead to privacy risks if smart contracts are not properly audited and secured. For example, outdated code, incorrect mathematical conditions, or poor design can result in unintended consequences and expose the contract to third-party exploitation.

It is important for developers to follow best practices for smart contract development, including thorough testing and auditing, to minimize these risks and ensure the security and privacy of smart contracts in permissionless blockchains. A well-known example of a smart contract vulnerability is the hack of the Decentralized Autonomous Organization (DAO) in 2016, where a hacker exploited a vulnerability in the smart contract code to steal a significant amount of funds. This incident highlights the importance of proper smart contract development and the need to continuously work towards improving the security and reliability of smart contracts in blockchain technology.

## IV. POTENTIAL SOLUTIONS

### A. Multi-signature Technique

A multi-sig wallet could help the participants who start to enter the permissionless blockchain protect their digital assets. more private keys from different sources help a participant reduce the privacy risks compared to single key management. Multi-signature (multi-sig) is a technique that can help address the privacy risks associated with public blockchains by requiring multiple parties to sign a transaction before it can be processed. Consider a scenario where three users, A, B, and C, hold private keys for a multi-sig address on a blockchain. For a transaction to be executed, a minimum of two of the three private keys must be used to sign it. For example, if A and B sign a transaction, it would be processed on the blockchain, but it is not possible to determine which of the two users signed the transaction.

While it can provide increased security and control over transactions, there are some disadvantages to consider such as complexity in implementation, slower transaction times, increased risk of failure if one party is unavailable or unwilling to sign, dependence on multiple parties which can lead to disputes or disagreements, and increased costs. It is important to weigh these potential drawbacks against the benefits before implementing multi-sig.

### B. Zero-Knowledge Proofs

Zero-knowledge proof (ZKP) is a promising solution to address privacy concerns in blockchain technology. ZKP allows one party to prove to another that a statement is true, without revealing any additional information beyond the statement itself. This makes ZKP an effective tool for maintaining privacy in blockchain transactions.

In a blockchain transaction, parties can use ZKP to prove that certain conditions have been met without revealing any details about the transaction or the parties involved. This allows for secure and private transactions on the blockchain $(P_1, P_2) \rightarrow (P_1|P_2)$, $P_1$ proves the validity of the transaction and $P_2$ verifies the proof without revealing any additional information.

$P_1$ represents the party initiating the transaction, and $P_2$ represents the party verifying the transaction. By using ZKP, the parties can ensure the privacy and security of the transaction without sacrificing transparency and trust. In Figure 2 presents the simplified framework of ZKP namely the prover and the verifier. The implementation of this framework consists of following three phases, witness phase, challenge phase and response phase, in the above phases, no private data would be leaked.

While ZKPs have the potential to provide increased security and privacy, there are also some disadvantages to consider. Implementing ZKPs can be complex and may require a high level of technical expertise, which can make it difficult for some users to effectively utilize the technology. Additionally, ZKPs can be computationally intensive and may result in slower transaction times, making them less efficient for some applications.

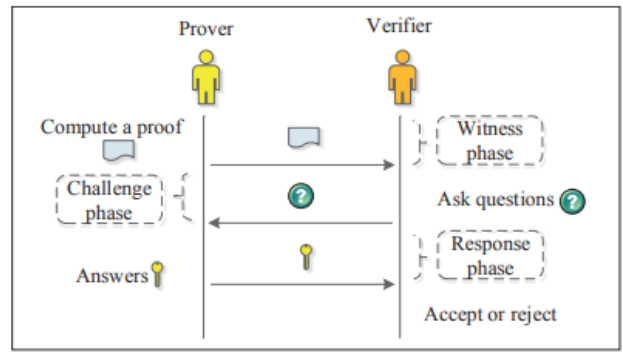

Figure 2. The framework of zero-knowledge proof.

This highlights the potential of ZKP as a solution for privacy issues in blockchain technology. Future research can explore the integration of ZKP with other privacy-preserving technologies to enhance the security and privacy of blockchain transactions.

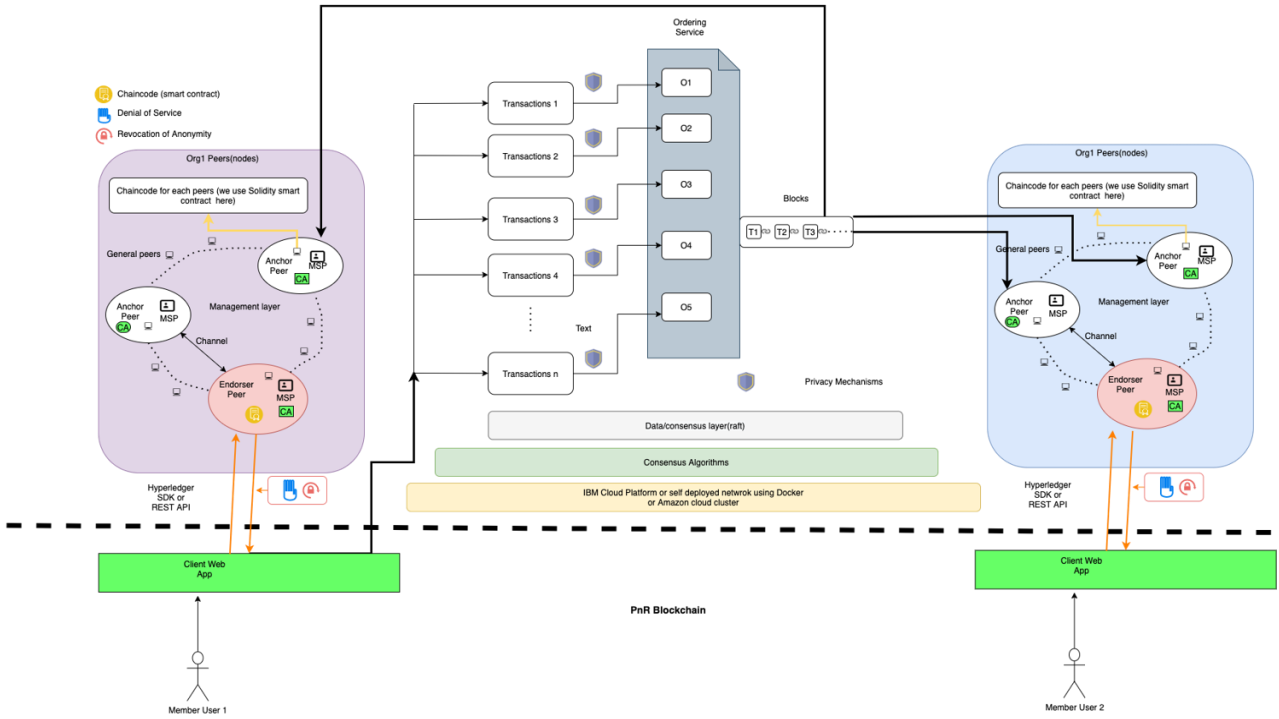

Figure 3. Test prototype of PnR Blockchain architecture.

## C. Hyperledger Fabric

Hyperledger Fabric addresses privacy concerns in public blockchain by implementing a permissioned network model, where only authorized entities have access to the ledger data. It also offers selective disclosure of sensitive information through channels, allowing only parties with a legitimate need to view certain transactions. This solution offers a more secure and controlled environment for businesses to transact on a blockchain network, ensuring the confidentiality of their data. In [9] author explained the fundament of the blockchain properties, analyzed potential threats at that time, and investigated the unique privacy requirements of permissionless blockchain.

In [1] author proposed a design of the Punishment not Reward (PnR) blockchain architecture. In Figure 3 we are prototyping a test network for the PnR architecture, we can see, the modular Hyperledger Blockchain offers a flexible solution for developing privacy-preserving blockchain networks.

## D. Ring Signatures

Ring signatures provide a solution to the privacy risks associated with public blockchains by allowing a group of users, say $U_1, U_2, ..., U_n$, to sign a message m without revealing the identity of the signer. A ring signature can be mathematically represented as $(m, U_1, U_2, ..., U_n, \sigma)$ where σ is a valid signature on message m under the public keys of the group $U_1, U_2, ..., U_n$ such that the identity of the signer is indistinguishable from the other members in the group. This makes it difficult for an observer to determine who among the group of users signed the message. By using ring signatures, public blockchains can offer increased privacy to users who wish to transact without revealing their identity. Additionally, ring signatures can also be used to implement anonymous voting and secure multi-party computations on public blockchains, making these systems more attractive for a wider range of use cases.

## E. Stealth Addresses privacy-enhancing technology

Stealth addresses, also known as one-time addresses, are a privacy-enhancing technique that can be used in blockchain technology. They are designed to protect the privacy of the recipient of a transaction by hiding their address. A stealth address is generated by the recipient, and the sender then uses this address to send the funds. The transaction is recorded on the blockchain, but the recipient's address remains private and cannot be linked to their identity.

Mathematically, a stealth address can be represented by a one-way function $H(x)$ where the recipient's public key $P$ is used as the input $x$. The sender can then use this function to generate a unique stealth address for each transaction. The recipient can then use their private key to access the funds at this address. We summarize and test the exact original method of (user 'bytecoin',2011) using the notations of slide 21 in (Courtois6,2016).

- The recipient has a public key.
- The sender uses public key.
- Now Diffie-Hellman allows both the sender and the recipient to compute a certain value S.

$$S = a.B = b.A \in E(\mathbb{F}\rho) \qquad (1)$$

- The ephemeral transfer address is then simply $H(s).G$ in $E(\mathbb{F}\rho)$, private key is $c = H(S) \bmod Q$ and in normal bitcoin operation only $H'(H(S).G)$ would be revealed initially when coins are sent to $pk = H(S).G$.

- The receiver actively monitors the blockchain or other channels for all plausible $A$ and checks if somebody is sending coins to some $H'(H(b.A).G)$. He can spend all such outputs.

In this way, stealth addresses provide an additional layer of privacy for blockchain transactions by obscuring the recipient's address and allowing them to receive funds without revealing their identity to the public. This can be especially important in cases where the recipient may wish to keep their financial activity private.

*F. zkEVM*

zkEVM is an implementation of Ethereum Virtual Machine (EVM) that uses zero-knowledge proofs to validate computations and ensure the privacy of smart contract execution. zkEVM enables smart contracts to run without revealing any confidential information to the Ethereum network. This provides a higher degree of privacy compared to traditional EVM and helps to mitigate privacy risks in public blockchains. By obscuring sensitive information, zkEVM can be used to protect user data, financial transactions, and other sensitive information stored on the blockchain.

## V. CONCLUSION

Privacy risks associated with permissionless blockchains are a major concern for individuals and businesses. The current solutions to these privacy risks, including zero-knowledge proofs, ring signatures, Hyperledger Fabric, zkEVM, and stealth addresses, provide some level of privacy but are not without their limitations. Future development of privacy solutions in the permissionless blockchain space should focus on improving the privacy and scalability of these solutions to meet the growing demand for privacy on permissionless blockchains.